# Resonantly enhanced second-harmonic generation using III-V semiconductor all-dielectric metasurfaces

Sheng Liu,*,†,‡ Michael B. Sinclair,† Sina Saravi,§ Gordon A. Keeler,† Yuanmu Yang,†,‡ John Reno,†,‡ Gregory M. Peake,† Frank Setzpfandt,§ Isabelle Staude,§ Thomas Pertsch,§ Igal Brener*,†,‡

†*Sandia National Laboratories, Albuquerque, New Mexico 87185, United States*

‡*Center for Integrated Nanotechnologies, Sandia National Laboratories, Albuquerque, New Mexico 87185, United States*

§*Institute of Applied Physics, Abbe Center of Photonics, Friedrich-Schiller-Universität Jena, Max-Wien-Platz 1, 07743 Jena, Germany*

*Corresponding authors: snliu@sandia.gov, ibrener@sandia.gov*

Abstract:

Nonlinear optical phenomena in nanostructured materials have been challenging our perceptions of nonlinear optical processes that have been explored since the invention of lasers. For example, the ability to control optical field confinement, enhancement, and scattering almost independently, allows nonlinear frequency conversion efficiencies to be enhanced by many orders of magnitude compared to bulk materials. Also, the subwavelength length scale renders phase matching issues irrelevant. Compared with plasmonic nanostructures, dielectric resonator metamaterials show great promise for enhanced nonlinear optical processes due to their larger mode volumes. Here, we present, for the first time, resonantly enhanced second-harmonic generation (SHG) using Gallium Arsenide (GaAs) based dielectric metasurfaces. Using arrays of cylindrical resonators we observe SHG enhancement factors as large as $10^4$ relative to unpatterned GaAs. At the magnetic dipole resonance we measure an absolute nonlinear conversion efficiency of $\sim 2 \times 10^{-5}$ with ~3.4 GW/cm$^2$ pump intensity. The polarization properties of the SHG reveal that both bulk and surface nonlinearities play important roles in the observed nonlinear process.

**Introduction:**

Nonlinear optical phenomena[1-3] were discovered immediately after the invention of lasers and have been widely used to broaden the spectral range accessible with lasers. In conventional nonlinear optical processes, obtaining high nonlinear optical conversion efficiency requires meeting strict phase-matching conditions of the interacting optical fields and this is achieved using uniaxial or biaxial bulk nonlinear crystals, or quasi-phase matching techniques[4]. Recently, advances in nanostructured optical materials, plasmonics, and metasurfaces have enabled nonlinear optical processes that do not depend on phase matching[5]. These approaches create tight confinement and large resonant enhancement of electromagnetic fields which generate much higher nonlinear efficiencies than in the constituent materials[5, 6]. For example, plasmonic nanoparticles and their assemblies have been intensively employed to study second- and third-harmonic generation[5, 7-12], and metasurfaces coupled to intersubband transitions have been used for second-harmonic generation (SHG) with efficiencies approaching those of macroscopically thick crystals[13, 14].

Recently, metasurfaces comprising arrays of Mie dielectric resonators have attracted much attention at optical frequencies due to their much lower loss compared with their metallic counterparts[15-17]. In particular, silicon has been used extensively as the constituent material for all-dielectric metamaterials that have been used for a variety of applications including high efficiency Huygens' metasurfaces[18, 19], beam steering[20], ultra-thin waveplates[21-23], zero-index directional emission[24] and polarization insensitive holograms[25]. In the last few years, it was realized that dielectric nanoresonators can also be used to greatly enhance nonlinear optical phenomena[26, 27], due to the largely enhanced electromagnetic fields inside the resonators and the larger mode volume. However, due to the centrosymmetric crystal structure of silicon, second-order nonlinear optical phenomena were not observed in Si-based metasurfaces. Therefore, dielectric metasurfaces based upon other materials that exhibit an intrinsic second order nonlinear susceptibility ($\chi^{(2)}$) are needed to fully exploit this approach for enhanced harmonic generation and other second-order nonlinear phenomena. Nanoscale resonators made from III-V semiconductors can fulfill these requirements.

Here we demonstrate, for the first time, resonantly enhanced SHG using dielectric metasurfaces that are made from GaAs which possesses a large intrinsic second-order nonlinearity of $d_{14}$~200 pm/V[4]. We study the second-harmonic (SH) response from GaAs nano-resonator arrays over a broad spectral range that encompasses both their electric and magnetic dipole resonances. At both resonances, we observe enhanced SHG that is orders of magnitude stronger than the SHG from unpatterned bulk GaAs. Most interestingly, the conversion efficiency at the magnetic dipole resonance is ~100 times higher than the conversion efficiency at the electric dipole resonance. This is in part due to the increased absorption of GaAs at the shorter wavelength of the electric dipole resonance. We also measured spectral dependence of the SHG polarization, from which we conclude that both bulk and surface nonlinearities play roles in the observed SHG response. Our investigations not only improve our understanding of nonlinear optical processes in these nanostructured materials, but also highlight the opportunities for nonlinear frequency up- and down-conversion without phase-matching, as well as entangled photon pair generation.

**Design and fabrication of GaAs dielectric metasurfaces**

We design the GaAs resonators to support Mie magnetic and electric dipole resonances at wavelengths longer than the GaAs bandgap to: 1) avoid absorption; and 2) to lie within the spectral range of our femtosecond Ti:sapphire laser. Note that, for the choice of dipole resonant frequencies used in this work, absorption will still occur at the SH wavelengths which are shorter than the GaAs bandgap. At the lowest dipole resonances, the resonators have side dimensions that are roughly $\lambda/n$, where $\lambda$ is the free space wavelength and n the refractive index. Our nonlinear metasurface comprises a square lattice of GaAs nanodisk resonators lying on a low refractive index $(Al_xGa_{1-x})_2O_3$ native oxide spacer layer that is formed by selectively oxidizing high-Al content $Al_xGa_{1-x}As$ layers[28-31]. The resonator array pitch is designed to minimize the interaction between neighboring GaAs resonators. Figure 1(a) shows the fabrication steps for creating the GaAs metasurfaces starting from molecular beam epitaxial growth of a 300-nm-thick layer of $Al_{0.85}Ga_{0.15}As$ followed by a 300-nm-thick layer of GaAs on top of a semi-insulating (100)-oriented GaAs substrate. We first spin-coat a negative tone hydrogen silsesquioxane (HSQ Fox-16) resist on the sample and pattern circular disks using standard electron-beam lithography that converts the HSQ to $SiO_x$. The unexposed HSQ is developed using Tetramethylammonium hydroxide leaving ~500-nm-tall $SiO_x$ nano-disks as etch masks for GaAs. The shape of the $SiO_x$ nano-disks is then transferred onto the GaAs and AlGaAs layers using an optimized chlorine-based inductively-coupled-plasma (ICP) etch recipe. Finally, the sample is placed in a tube furnace at ~420 degrees Celsius for a selective wet oxidization process that converts the layers of $Al_{0.85}Ga_{0.15}As$ into its oxide $(Al_xGa_{1-x})_2O_3$ which has a low refractive index of n~1.6[28]. The large refractive index contrast between the GaAs resonators and the underlying oxide ensures well defined Mie modes with tightly-confined electromagnetic fields inside the resonators — all essential for efficient nonlinear optical generation. Note that this fabrication is simplified from our previous work that used extra processing steps to create a $SiO_2$ etch mask for GaAs[32].

Figure 1(b) & (c) show 75-degree side-view and top-view SEM images of a metasurface consisting of an array of GaAs resonators with diameters of ~250 nm and heights of 300 nm. The side view image shows clear color contrast between the top $SiO_x$ etch mask, the GaAs resonators in the middle, and the AlGaO nano-disks at the bottom. The etch masks are not removed since, due to the low refractive index of $SiO_x$, they barely perturb the distribution or intensity of the electromagnetic fields within the GaAs resonators. The sample has an array pitch of 600 nm resulting in a spacing of ~350 nm between resonators so the interaction between the neighboring resonators is negligible.

We measured the linear-optical reflectivity spectrum (inset of Figure 1(c)) of the GaAs dielectric metasurface using a custom-built white-light spectroscopy setup. We used a 20X Mitutoyo Plan Apo NIR infinity-corrected objective (numerical aperture = 0.4) to both focus a broadband light source onto the samples and collect the reflected light. The reflected light was dispersed using a spectrometer with appropriate gratings and detected using a liquid nitrogen cooled InGaAs camera. The measured reflectivity spectrum was then normalized by the spectrum of a gold mirror measured under the same conditions. The normalized spectra exhibit well-separated magnetic and electric dipole resonances at ~1020 nm and 890 nm, respectively, thereby confirming the formation of the low loss GaAs dielectric metasurface. The high reflectivity (~100%) of the metasurface at the magnetic dipole resonance is due to the low loss of high-quality crystalline GaAs below the bandgap.

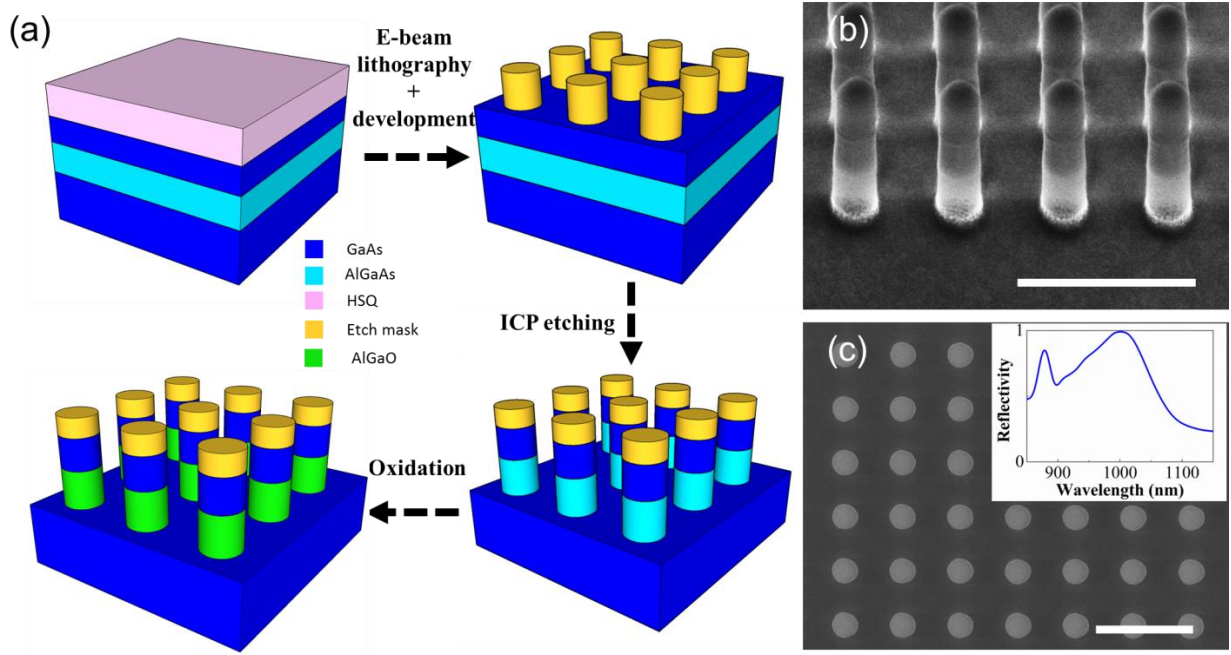

**Figure 1.** (a) Steps for fabricating GaAs based dielectric metasurfaces starting with molecular beam epitaxial growth, followed by e-beam lithography patterning, ICP dry etching, and selective wet oxidation. (b) 75-degree side view and (c) top view SEM images of the fabricated GaAs dielectric resonator array. The GaAs resonators have the same diameter of ~250 nm and height of 300 nm. The inset of (c) is the reflectivity spectrum of the GaAs resonator array which exhibits two well-separated reflectivity peaks corresponding to the magnetic and electric dipole resonances. The scale bars correspond to 1 μm.

**Resonantly enhanced SHG in GaAs resonators**

GaAs is known to possess large second-order nonlinearities with $d_{14}$=~200pm/V. This value is much higher than in conventional nonlinear crystals such as β-Barium Borate ($d_{22}$~2.2pm/V) and LiNbO$_3$ ($d_{31}$=~6pm/V and $d_{33}$=~30pm/V)[4, 33]. However, efficient SHG using GaAs has been challenging due to the difficulty in meeting phase-matching conditions for long crystals in the zinc-blende crystal structure which exhibits isotropic refractive indices. In addition, (100)-GaAs possesses only one non-zero $\chi^{(2)}$ tensor element ($d_{14}$) which restricts the choice of nonlinear optical device geometries. In the following we will show that, due to the subwavelength layer thicknesses, resonantly enhanced SHG can be obtained from our GaAs dielectric metasurfaces without any provision for phase matching.

We performed the SHG measurements in reflection geometry because the SHG wavelengths are above the bandgap of GaAs so that the SH signal in the transmission direction would be completely absorbed by the GaAs substrate. Figure 2(a) shows the experimental setup for measuring reflected SHG intensities and polarizations. We define the coordinate axes as shown in the figure: the sample surface is the x-y plane, and pump propagates along the z axis. Our optical pump was a mode-locked tunable Ti:sapphire laser oscillator that produced horizontally polarized pulses with 80-MHz repetition-rate and ~120 *f*s pulse width. The pump beam was reflected by a dichroic beam splitter and then focused to a spot diameter of ~6 μm on the sample using a 20X near-infrared objective. The generated SH was collected by the same objective, then transmitted through the beam splitter and measured using either a power meter or a near-IR spectrometer. The polarization of the SHG was measured using a linear polarizer. The detection efficiency of the whole system was calibrated using a broadband calibration lamp. To simplify the physical interpretation, we rotated the sample about the z-axis so that the pump polarization (along the x axis) was parallel to the [010]-direction of the GaAs wafer.

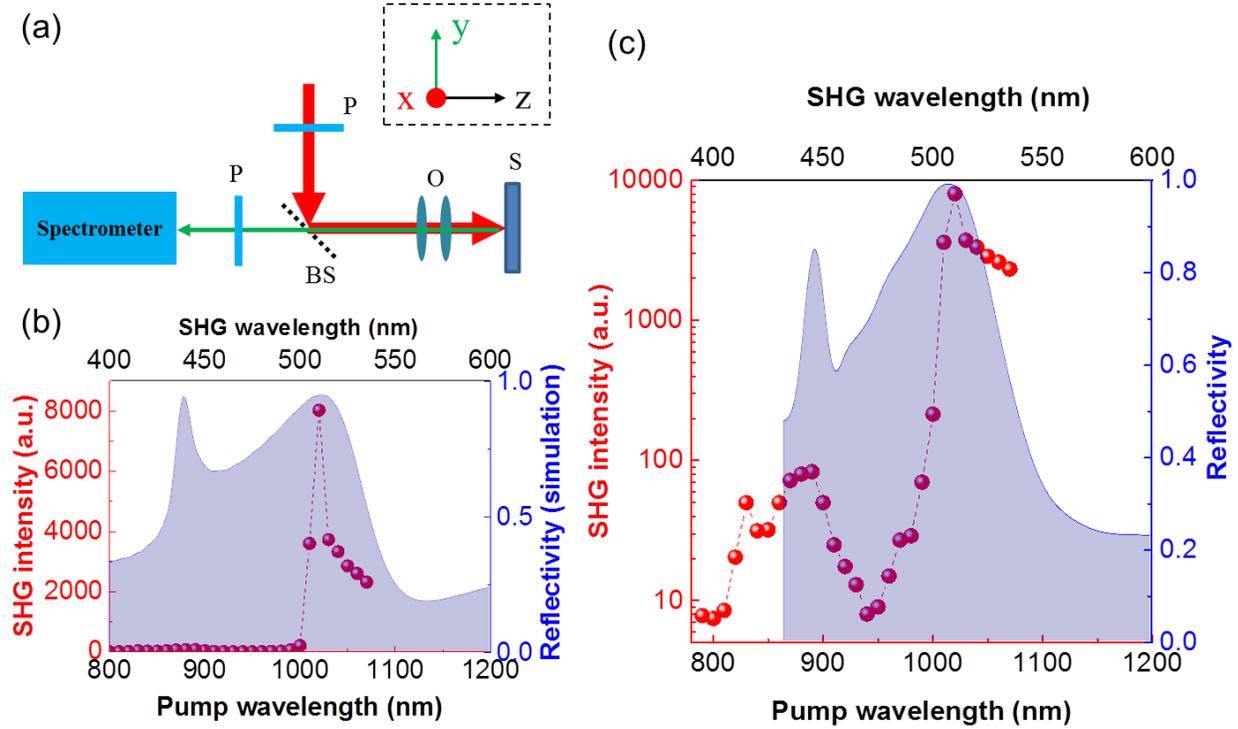

**Figure 2.** (a) Experimental setup for SHG measurements in the reflection geometry. (S: sample, O: objective, P: polarizer, BS: beam splitter). The inset shows that the pump propagates along the z axis and the pump polarization is along the x axis. Experimental results for the spectral dependence of the SHG intensity showing resonantly enhanced SHG behavior at the magnetic and electric dipole resonances on (b) linear and (c) logarithm scales. The (b) simulated and (c) experimental linear reflectivity spectra of the sample are used as the backgrounds.

Figures 2(b) & (c) show the SHG intensity on linear and logarithmic scales, respectively, as the pump wavelength is tuned while keeping the pump power constant. The simulated and experimental linear reflectivity spectra are used as the backgrounds for (b) & (c), respectively. The SHG power exhibits peaks in the vicinity of the magnetic (~1020 nm) and electric (~890 nm) dipole resonances due the electromagnetic field enhancements that occur at these resonances. Indeed, electromagnetic simulations described below show that at these resonances the field

intensities are ~30 times stronger than the incident pump intensity. The SH signal obtained when the pump coincides with the magnetic and electric dipole resonances are more than 3 and 1 orders of magnitude higher than the signal obtained when pumping at off-resonant wavelengths. We believe that the large difference in SHG intensities obtained at the magnetic and electric resonances is partly due to higher absorption of GaAs at the shorter SH wavelength associated with the electric dipole resonance, and partly due to different origins of the SHG response at the two dipole resonances, as will be discussed below. We note that the peak SH signals at the two dipole resonances are even larger (~4 orders of magnitude at the magnetic dipole resonance) than the SH signal obtained from unpatterned GaAs. Therefore, SHG at off-resonant wavelengths is still ~10 times stronger than that generated at unpatterned regions. This is likely contributed by the enhancement of the electric field inside the GaAs resonators even at off-resonant wavelengths (see Supporting Information, Section 1). Note that all the experiments were performed under the same optical excitation conditions; we only changed the pump location on the sample. Also note that a small z-polarized electric field component was introduced by the finite NA of the focusing objective that we used (NA ~ 0.42).

The power dependence of the SHG signal is shown in Figure 3(a). This measurement was performed for a pump wavelength of 1020 nm which corresponds to maximum SHG efficiency as shown in Figure 2. The quadratic power relationship is maintained over a wide pump power range until irreversible damage of the GaAs resonators occurred at an average power of ~5 mW (peak intensity of ~1.5 GW/cm$^2$) as shown in the inset of Figure 3(a). At ~11 mW average pump power excitation, the SHG power continuously decreased over time (the black triangles) due to the physical damage to the sample. The inset of Figure 3(b) shows the severe damage caused to the GaAs resonators after illumination by a much higher average power of 27 mW (peak intensity of ~8.1 GW/cm$^2$). This damage was likely associated with two-photon-absorption of GaAs followed by thermal damage due increased free carrier absorption enhanced by the high electric field intensity inside the resonators. Therefore, scaling to higher pump powers would require the fabrication of larger resonators so the dipole resonances (and, hence the pump photon energy) can be tuned to below half of the GaAs bandgap. Note that surface defects created during the process of ICP dry etch could increase the loss of GaAs and therefore contribute to the damage. Figure 3(a) also shows a ~4 orders of magnitude enhancement of SHG from the GaAs metasurface compared with SHG on unpatterned GaAs.

Figure 3(b) shows that the SHG conversion efficiency increases as the pump power increases and reaches a maximum conversion efficiency of $\sim 2 \times 10^{-5}$ when the pump power is ~11.4 mW (peak intensity of ~3.4 GW/cm$^2$). Before reaching the damage threshold of ~5 mW, the nonlinear coefficient is $\sim 1.5 \times 10^{-8} W/W^2$, which is ~two orders of magnitude higher than recently published record high SHG efficiency using mode-matching plasmonic nanoantennas[9]. This increase is even more impressive, considering that the pump beam was much more tightly focused in Ref. 9, while the nonlinear coefficient (which is defined solely in terms of powers) scales inversely with the spot size. Furthermore, a higher NA objective (1.35) was used in Ref. 9, as compared to the NA =0.42 objective used in the current work which allowed collection of a larger portion of the radiated SHG power (some of which is emitted in diffraction lobes).

The SHG enhancements arising from the electromagnetic field enhancements of the GaAs metasurfaces can be treated using the effective second-order nonlinear susceptibility tensor of the metasurface[14]:

$$\chi_{ijk}^{(2)\text{eff}} = \frac{\chi_{mnp}^{(2)}}{V} \int_V f_{m(i)}^{2\omega}(x,y,z) f_{n(j)}^{\omega}(x,y,z) f_{p(k)}^{\omega}(x,y,z) \mathrm{d}V$$

where $\chi_{mnp}^{(2)}$ is the material's intrinsic second-order nonlinear susceptibility, $V$ is volume, $f^{2\omega}$ is the field enhancement at the SH wavelength, and $f^{\omega}$ is the field enhancement at the fundamental wavelength. In this case, the SHG power is proportional to $\chi_{ijk}^{(2)\text{eff}} \cdot I_p^2$, where $I_p$ is the incident pump intensity. Therefore, it is important to achieve high electromagnetic field enhancements at both the fundamental and SH wavelengths. However, our simulations show weak electromagnetic fields inside the resonators at the SH wavelengths due to the large absorption of GaAs at visible wavelengths which limits our SHG efficiency (see Supporting Information, Section 2). Simulations also show that the electromagnetic fields are much weaker at the SH of the electric dipole wavelength than at the SH of the magnetic dipole wavelength, which partly explains the large difference between the SHG powers obtained when pumping at the two dipole resonances. Therefore, we expect that large improvements of the SHG conversion efficiency could be obtained by designing larger GaAs resonators so that the magnetic dipole resonance occurs at wavelengths longer than twice of the GaAs bandgap wavelength or by replacing the GaAs with AlGaAs which has higher bandgap energy. This would: 1) minimize absorption at the SH wavelength, thereby allowing for significant electromagnetic field enhancements at the SH wavelength (see Supporting Information, Section 3); and 2) minimize two-photon-absorption induced damage at larger pump powers. Further improvements could also be obtained by optimizing the resonator shape in order to obtain a maximum modal overlap between the SH and fundamental wavelengths[9].

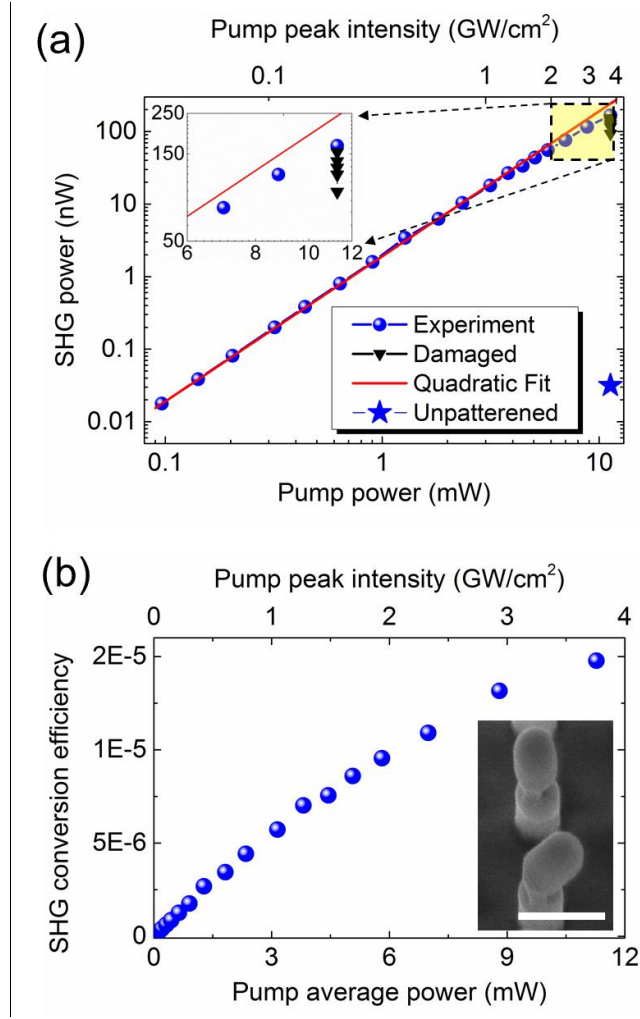

**Figure 3.** (a) Quadratic relationship between the average pump and SHG powers at low pump intensities and the deviation from the quadratic relationship at higher pump intensities due to the damage of GaAs resonators. (b) SHG conversion efficiency as a function of the pump power. The inset of (b) is an SEM image of damaged GaAs resonators resulting from illumination at a high average pump power of ~27 mW.

Studies of the SHG polarization have previously been used to elucidate the nonlinear optical processes in nanostructured materials[8, 9, 11, 12, 34, 35]. Due to the large surface to volume ratio in semiconductor based nanostructures, the SHG contributed by surface nonlinearities is often comparable to or even larger than that from bulk nonlinearities[36-38]. Using full-wave time-dependent simulations including second order nonlinearities we first simulated the SHG emitted from our GaAs metasurfaces assuming only a bulk nonlinearity tensor (see Supporting Information, Section 4 for more details). The bulk second-order susceptibility $\chi_{ijk}^{(2)}$ of GaAs is non-zero only when $i{\neq}j{\neq}k$ due to the zinc-blend crystal structure ($\bar{4}3m$) of GaAs[4]. Therefore, the possible generated nonlinear polarizations at the SH frequencies inside the GaAs nano-resonators are $P_{NL,x}^{2\omega} \propto 2\chi_{xyz}^{(2)} E_y^\omega E_z^\omega$, $P_{NL,y}^{2\omega} \propto 2\chi_{yxz}^{(2)} E_x^\omega E_z^\omega$ and $P_{NL,z}^{2\omega} \propto 2\chi_{zxy}^{(2)} E_x^\omega E_y^\omega$. We performed our nonlinear simulation using a freely available finite difference time-domain software package[39] and the details can be found in Supporting Information (Section 4). Although for an x-axis polarized and normally incident pump, $E_x^\omega$ and $E_z^\omega$ are the main components of the fundamental field inside the resonators (see Supporting Information, Section 5), we performed our simulation considering the contributions of all three electric field components for higher accuracy. Note that the simulation was performed with normally incident pump on the sample without considering the oblique angle incidence caused by the objective.

Figure 4(a) shows the calculated spectrally dependent SHG intensity with clear resonant enhancements at the magnetic and electric dipole resonances[4]. The spectral dependence of relative SHG intensity also agrees well with our experimental results shown in Figure 2(c). These results also agree with a recent theoretical work that attributed the SHG in AlGaAs nanoantennas to only bulk nonlinearity[40]. However, this nonlinearity assignment should be further verified by measuring the SHG polarization[36-38]. The polarization study is of particular interest considering the different field profiles at the electric and magnetic dipole resonances as Figures 4 (b) & (c) show. Figures 4(d) & (f) show the simulated polar plots of the SH signal polarization when pumping at the electric and magnetic dipole resonances, respectively. The maximum SH intensities at the two resonances are obtained along two orthogonal directions (x-and y-axis, using the same coordinate system as shown in Figure 2(a)). This is drastically different from our experimentally measured polar plot shown in Figure 4 (e) & (g) where the SH signal is polarized mainly along y- and x-axis at the electric and magnetic resonances, respectively. Note that the limited angular collection efficiency of our objective (NA~0.42) was taken into account for the polar plot calculation. These experimental findings rule out the bulk nonlinearity as the sole source of the observed SHG. Instead, we consider the surface nonlinearities, since at the GaAs surface, the bulk crystal symmetry $\bar{4}3m$ is broken, and the surface symmetry is $mm2$[4]. At the surface, the only nonzero nonlinear tensor components are $\chi_{xzx}$, $\chi_{xxz}$, $\chi_{yyz}$, $\chi_{yzy}$, $\chi_{zxx}$, $\chi_{zyy}$, $\chi_{zzz}$, assuming the surface normal direction is along the z-axis. For the simulation of SHG at the surface, we only consider the $\chi_{zzz}$ contribution due to its much higher value compared with the other components[36]. Moreover, due to the complex structure of the resonator sidewalls that have rotating normal directions, here we only simulated SHG from the top and bottom surfaces of the resonator, which have surface normal along z-axis. The resonantly enhanced SHG is also observed around both dipole resonances (see Supporting Information, Section 6). Figure 4(h) shows that when the pump is tuned to the electric dipole resonance, the simulated polar plot exhibits maximum intensity polarized along the y-axis, agreeing with the experimental results. At the magnetic dipole resonance, when we incorporate both the bulk and surface nonlinearities, we observe the rotation of the simulated SHG polarization observed in the far field due to the optically coherent interference between the two nonlinear emissions (see Supporting Information, Section 7). A direct comparison between the absolute SH intensity originating from surface and bulk nonlinearities remains challenging due to the unexplored value of the surface nonlinearity coefficient[36]. Moreover, further efforts are needed to obtain a comprehensive understanding by considering the sidewall's contribution as well as the impact of non-normal incidence that causes the redistribution of electric field enhancement inside the resonators. Nonetheless, we show that the surface mediated SHG can exhibit very different behavior than the SHG due to a bulk nonlinearity. Moreover, as has been demonstrated in other materials[41-44], lattice mismatch between the GaAs and AlGaO underlayer induces strain in the GaAs close to the AlGaO interface which may also change the surface nonlinearity. Finally, figure 4(i) shows the SH spectra obtained as the location of the focused pump beam is varied across the metasurface. Although the peak of the SH signal always occurs in the vicinity of half the pump wavelength, we observe slight spectral shifts with the pump beam position. We attribute this to slight dimensional non-uniformities across the array which causes the actual wavelength of peak field enhancement to vary with position.

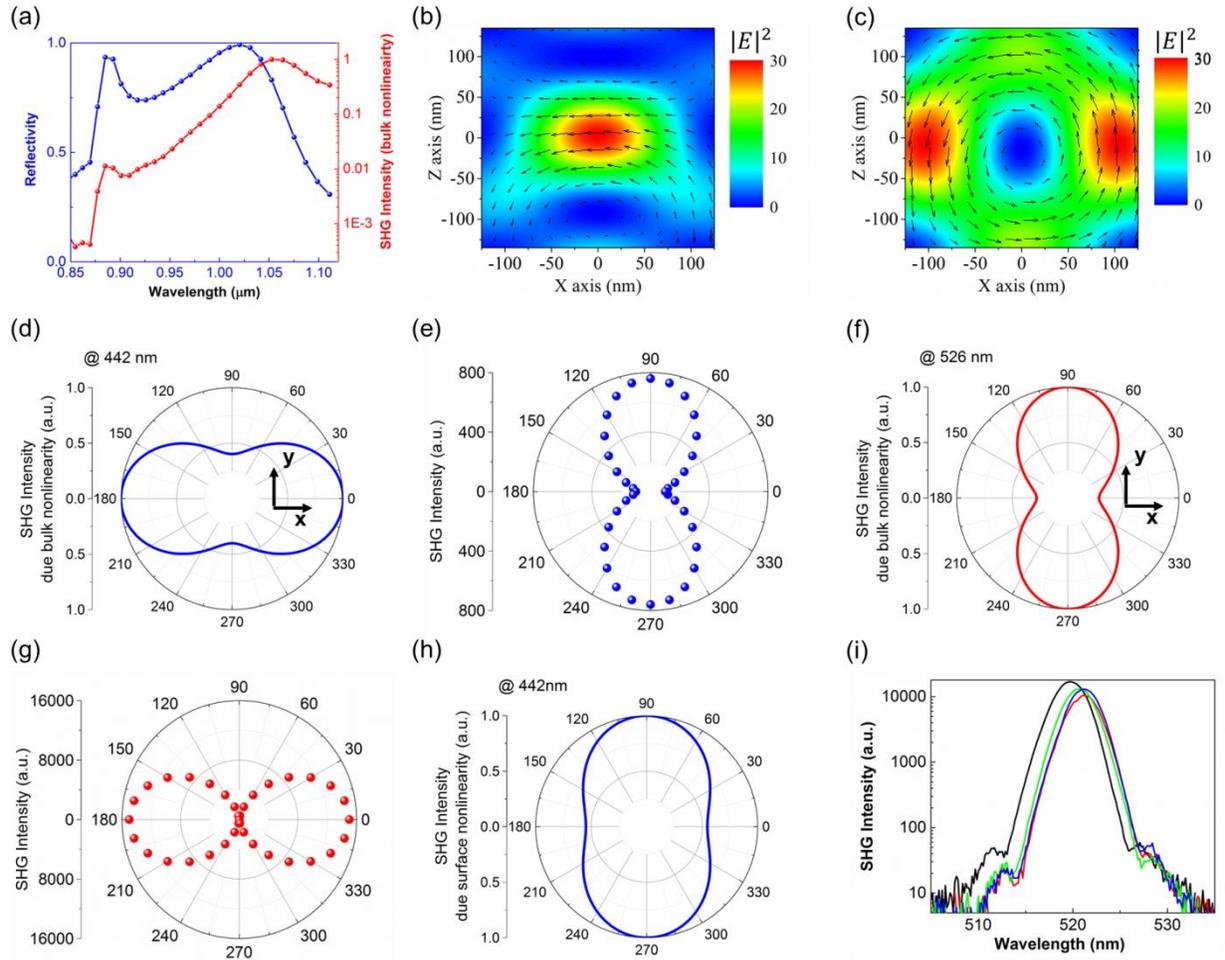

**Figure 4.** (a) Spectral dependence of the SHG intensity originating from a bulk nonlinearity calculated using full-wave simulations. Simulated electric intensity profile at the (b) electric and (c) magnetic dipole resonances in the x-z plane located half way through the GaAs resonator. The edges of (b) and (c) are the boundaries of the GaAs nanoresonator. Polar plots of the SH polarization obtained by simulation using bulk nonlinearity ((d) and (f)) and by experimental measurements ((e) and (g)) when pumping with an x-polarized beam at the electric ((d) and (e)) and magnetic ((f) and (g)) dipole resonances. In these plots, 0 and 90 degrees represent polarizations along the x and y axis, respectively. (h) Simulated polar plot of SH polarization due to the surface nonlinearity from the top and bottom GaAs resonator surfaces when pumping

at the electric dipole resonance; the polarization of the SHG signal now agrees with the experimental observations in (e). (i) Slight variation of SH spectra due to the fabrication imperfection caused non-uniformities of GaAs resonators' dimensions.

**Conclusions**

In summary, we have experimentally demonstrated resonantly enhanced SHG using GaAs dielectric metasurfaces fabricated from (100)-oriented GaAs. Our nonlinear coefficient of ~$1.5 \times 10^{-8}$ W/W$^2$ is much higher than the record high SHG obtained using mode-matching of plasmonic nanoantennas[9]. We attribute this increase to the resonantly enhanced electromagnetic fields and the larger mode volume of the dielectric resonator; and to the larger nonlinear susceptibility of GaAs. We anticipate that even higher conversion efficiencies could be obtained by operating at longer wavelengths where GaAs absorption at the SH wavelength can be reduced allowing for SH field enhancements within the resonator, and by optimizing the overlap between the resonator modes at the SH and fundamental wavelengths. The experimentally observed polarization dependence of the SH signal indicates that bulk nonlinearities are not the sole source of the observed SHG. While further investigations are needed to fully understand the SH polarization behavior, we showed that surface nonlinearities may contribute significantly to the observed SHG as well as affect the polarization of the measured SH in the far field. Our demonstration paves the way for using dielectric metasurfaces in other phase-matching free nonlinear optical applications such as next-generation nonlinear optical convertors for frequency mixing, photo pair generation, and all-optical-optical control and tunability.

**Acknowledgement**

Parts of this work were supported by the U.S. Department of Energy, Office of Basic Energy Sciences, Division of Materials Sciences and Engineering and performed, in part, at the Center for Integrated Nanotechnologies, an Office of Science User Facility operated for the U.S. Department of Energy (DOE) Office of Science. Sandia National Laboratories is a multi-program laboratory managed and operated by Sandia Corporation, a wholly owned subsidiary of Lockheed Martin Corporation, for the U.S. Department of Energy's National Nuclear Security Administration under contract DE-AC04-94AL85000.

# Supporting information

**Resonantly enhanced second-harmonic generation using III-V semiconductor all-dielectric metasurfaces**

Sheng Liu,*,[†,‡] Michael B. Sinclair,[†] Sina Saravi,[§] Gordon A. Keeler,[†] Yuanmu Yang,[†,‡] John Reno,[†,‡] Gregory M. Peake,[†] Frank Setzpfandt,[§] Isabelle Staude,[§] Thomas Pertsch,[§] Igal Brener*,[†,‡]

[†]*Sandia National Laboratories, Albuquerque, New Mexico 87185, United States*

[‡]*Center for Integrated Nanotechnologies, Sandia National Laboratories, Albuquerque, New Mexico 87185, United States*

[§]*Institute of Applied Physics, Abbe Center of Photonics, Friedrich-Schiller-Universität Jena, Max-Wien-Platz 1, 07743 Jena, Germany*

*Corresponding authors: snliu@sandia.gov, ibrener@sandia.gov

**S1: Electric field enhancement at an off-resonant wavelength of 965 nm.**

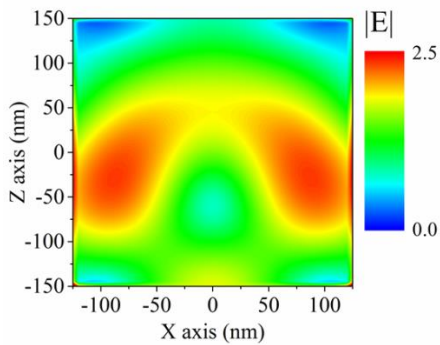

Figure S1. Simulated electric field ($|E|$) profiles at an off-resonant wavelength of 965 nm in the x-z plane located half way through the GaAs resonator.

**S2: Weak electromagnetic fields inside the resonators at the SH wavelengths due to the strong absorption of GaAs at visible frequencies (resonators' dimension: diameter 250 nm and height 300 nm).**

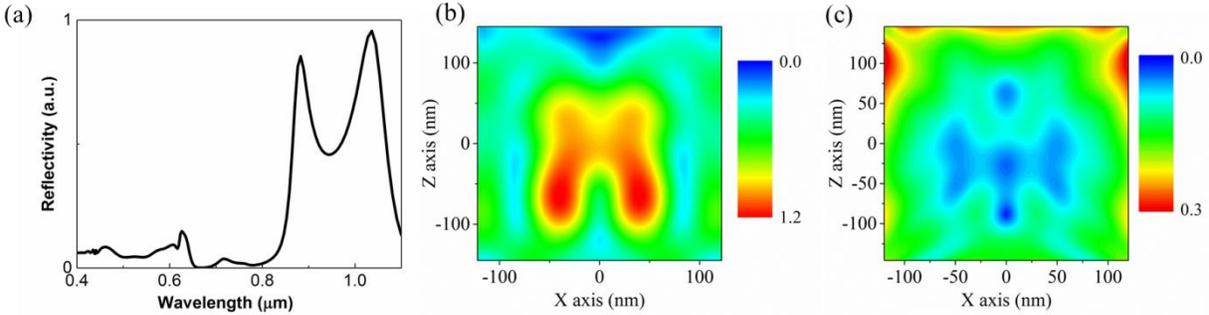

Figure S2. (a) Simulated reflectivity spectrum of GaAs dielectric resonator (D=250 nm, H=300 nm) arrays. Simulated electric field ($|E|$) profiles at the SH of (b) magnetic and (c) electric dipole resonances in the x-z plane located half way through the GaAs resonator. Electromagnetic fields are weak inside the resonators due to the absorption of GaAs at these short wavelengths. The weaker electric field at the SH of electric dipole resonance also partly contributes to the weaker SHG intensity at the electric dipole resonance.

**S3: Large electromagnetic field enhancements at the SH wavelengths inside larger GaAs resonators (resonators' dimension: diameter 600 nm and height 600 nm).**

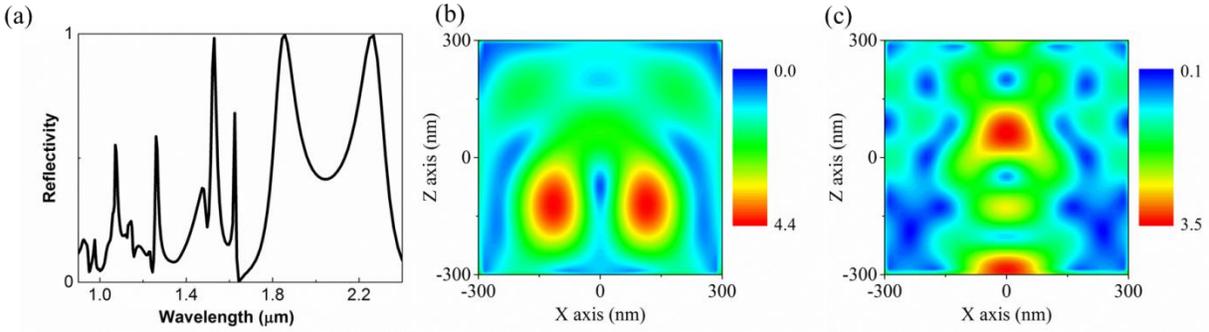

Figure S3. (a) Simulated reflectivity spectrum of GaAs dielectric resonator (D=600 nm, H=600 nm) arrays. Due to the larger dimensions, the SH of the electric and magnetic dipole resonances are below the bandgap of GaAs. Simulated electric field ($|E|$) profiles at the SH of (b) magnetic and (c) electric dipole resonances in the x-z plane located half way through the GaAs resonator. Electromagnetic fields are enhanced inside the resonators at these two SH wavelengths.

**S4: Details of the SHG nonlinear simulation.**

We performed our simulation by first retrieving all three electric field components at the fundamental frequencies inside the resonators. Next, the electric fields were multiplied using these three different tensor/field combinations for each location inside the resonators to generate the nonlinear polarizations. Finally, we placed these nonlinear polarizations back inside the resonators as sources at the SH frequencies.

**S5: Electromagnetic field profiles inside the GaAs resonators at the electric and magnetic dipole resonances.**

The Ex and Ez are the main components inside the GaAs resonators, especially in the x-z plane located half way through the GaAs resonator. Ey components is zero in the x-z plane located half way through the GaAs resonator, but not negligible when the plane move aside from the center of the resonators.

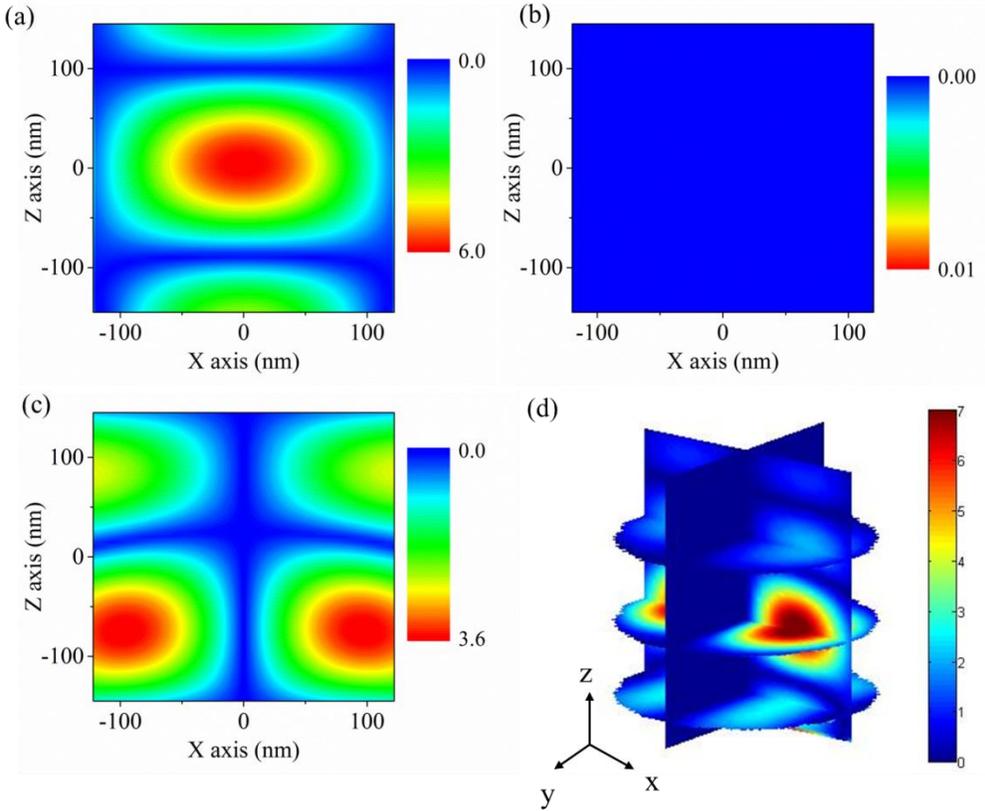

Figure S5. At the electric dipole resonance, the simulated electric field ($|E|$) profiles with polarization along (a) x axis, (b) y axis, and (c) z axis in the x-z plane located half way through the GaAs resonator. Electromagnetic fields polarized along y axis are negligible at this particular plane but not negligible at other locations inside the resonators. (d) The simulated field profiles of $|E_x^\omega \times E_z^\omega|$.

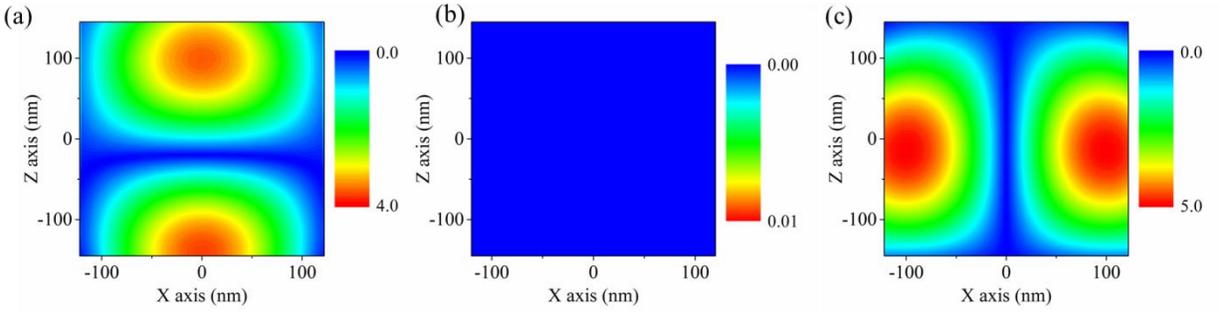

Figure S6. At the magnetic dipole resonance, the simulated electric field ($|E|$) profiles with polarization along (a) x axis, (b) y axis, and (c) z axis in the x-z plane located half way through the GaAs resonator. Electromagnetic fields polarized along y axis are negligible.

**S6: Spectrally dependent SHG intensity due to the surface nonlinearity from resonators' top and bottom surfaces.**

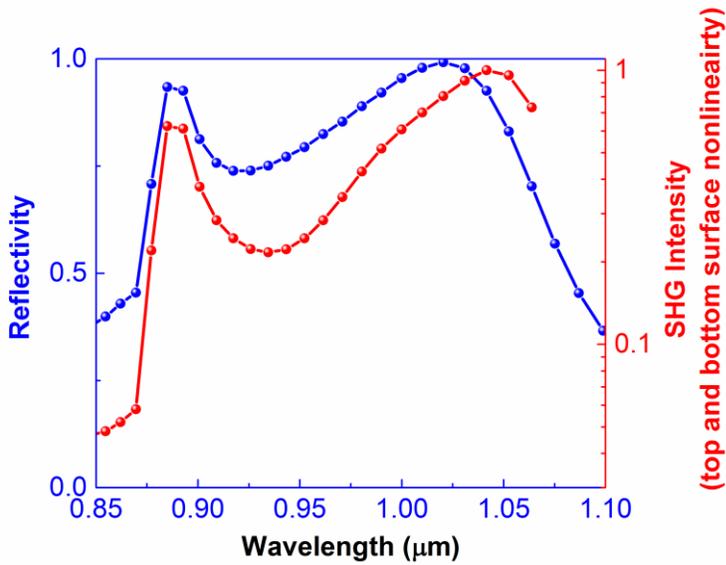

Figure S7. Simulation results of SH intensity from GaAs resonators' top and bottom surfaces showing resonant enhancement at electric and magnetic dipoles.

**S7: Rotation of the SHG polarization due to the interference between surface and bulk nonlinearities.**

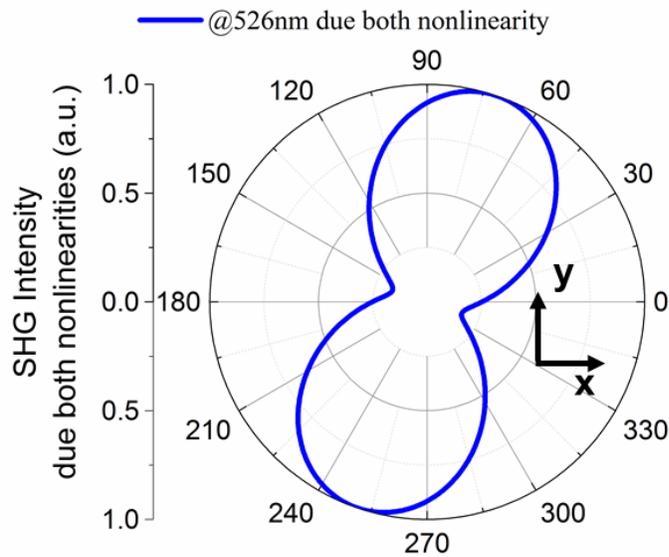

Figure S8. Full-wave simulation of the SH polarization projected to the far-field, when the pump is tuned to the magnetic-dipole resonance. This simulation uses both bulk and surface nonlinearities; for the later, only the top and bottom surfaces of the GaAs resonators were considered. Even though both bulk and surface nonlinearities cause maximum nonlinear polarizations along the y-axis, the simulated SH polarization rotates away from the y-axis, due to the interference between the two nonlinear emissions.